\documentclass[fleqn,usenatbib]{mnras}

\usepackage[T1]{fontenc}

\usepackage{graphicx}	
\usepackage{amsmath}	
\usepackage{amssymb}	
\usepackage{newtxtext,newtxmath}

\title[Large-scale expansion of OB stars in Cygnus]{Large-scale expansion of OB stars in Cygnus}
\author[A. L. Quintana and N.J. Wright]{
Alexis L. Quintana$^{1}$\thanks{E-mail: a.l.p.quintana.isasi@keele.ac.uk} and
Nicholas J. Wright$^{1}$ \\
$^{1}$Astrophysics Group, Keele University, Keele ST5 5BG, UK\\
}

\date{Accepted 2022 May 30. Received 2022 May 24; in original form 2022 March 23}

\pubyear{2022}

\begin{document}
\label{firstpage}
\pagerange{\pageref{firstpage}--\pageref{lastpage}}
\maketitle

\begin{abstract}
The proper motions (PMs) of OB stars in Cygnus have recently been found to exhibit two large-scale kinematic patterns suggestive of expansion. We perform a 3D traceback on these OB stars, the newly-identified OB associations and related open clusters in the region. We find that there are two groups of stars, associations and clusters and that they were each more compact in the past, reaching their closest approach $7.9^{+3.0}_{-1.8}$ and $8.5^{+0.8}_{-2.8}$ Myr ago. We consider two main scenarios for the driver of these large-scale expansion patterns: feedback-driven expansion from a previous generation of massive stars, and expansion as a result of the turbulent velocity field in the primordial molecular cloud. While it is tempting to attribute such large-scale expansion patterns to feedback processes, we find that the observed kinematics are fully consistent with the turbulent origin, and therefore that the injection of further energy or momentum from feedback is not required. Similar conclusions may be drawn for other star forming regions with large-scale expansion patterns. 
\end{abstract}




\section{Introduction}

Many OB stars are found in groups known as OB associations \citep{McKee}, which are characterized by their unbound nature and their low-density \citep[< 0.1 $M_{\odot}$ pc$^{-3}$,][]{Amb1947,Wright2020}. They are thought to constitute an intermediate step between star-forming regions, many of which are observed to be expanding \citep{Wright2019}, and the field population of stars. They are therefore useful to understand the star formation process and the build-up of the Galactic field \citep[e.g.,][]{Armstrong2020}.

Following star formation, stars, star clusters and OB associations disperse from their birth place due to various processes. This can include the intrinsic dispersion of velocities that the stars are born with, two- or three-body interactions between stars in clusters, and changes in the local gravitational potential (e.g., due to residual gas expulsion, \citealt{Lada2003}).

The intrinsic velocity dispersion within molecular clouds sets the initial kinematics of stars and is likely to dominate their kinematics on large (10--100 pc) scales. This is commonly observed in the form of a  power-law relationship between the linewidth (1D velocity dispersion) and the size of the molecular cloud. This was first observed by \citet{Larson} and since observed and refined by many studies (e.g., \citealt{Solomon, Bolatto, Miville}). This relationship is attributed to turbulence, which itself is driven by gravitational instabilities in the disk of the galaxy, magnetorotational instabilities and stellar feedback \citep{Miville}.

Feedback may also drive large-scale expansion patterns as energy and momentum is injected into the molecular cloud from a previous generation of massive stars. This feedback may be a combination of photoionization, radiation pressure, stellar winds, outflows and supernova explosions \citep{Dale2015}. Feedback is thought to be the process responsible for disrupting star formation and dispersing giant molecular clouds \citep{DobbsPringle2013}. It also injects large amounts of kinetic energy and momentum into the ISM \citep{Geen, Kim2015, Walch} which may introduce large-scale motions that are imprinted in the kinematics of future generations of stars. Astrometry from Gaia, especially its most recent data release \citep{GaiaEDR3}, can be very powerful to identify such kinematic patterns, as recent studies highlight \citep{Kounkel2020, Drew2021, Grob}.

\begin{figure*}
    \centering
    \includegraphics[scale = 0.33]{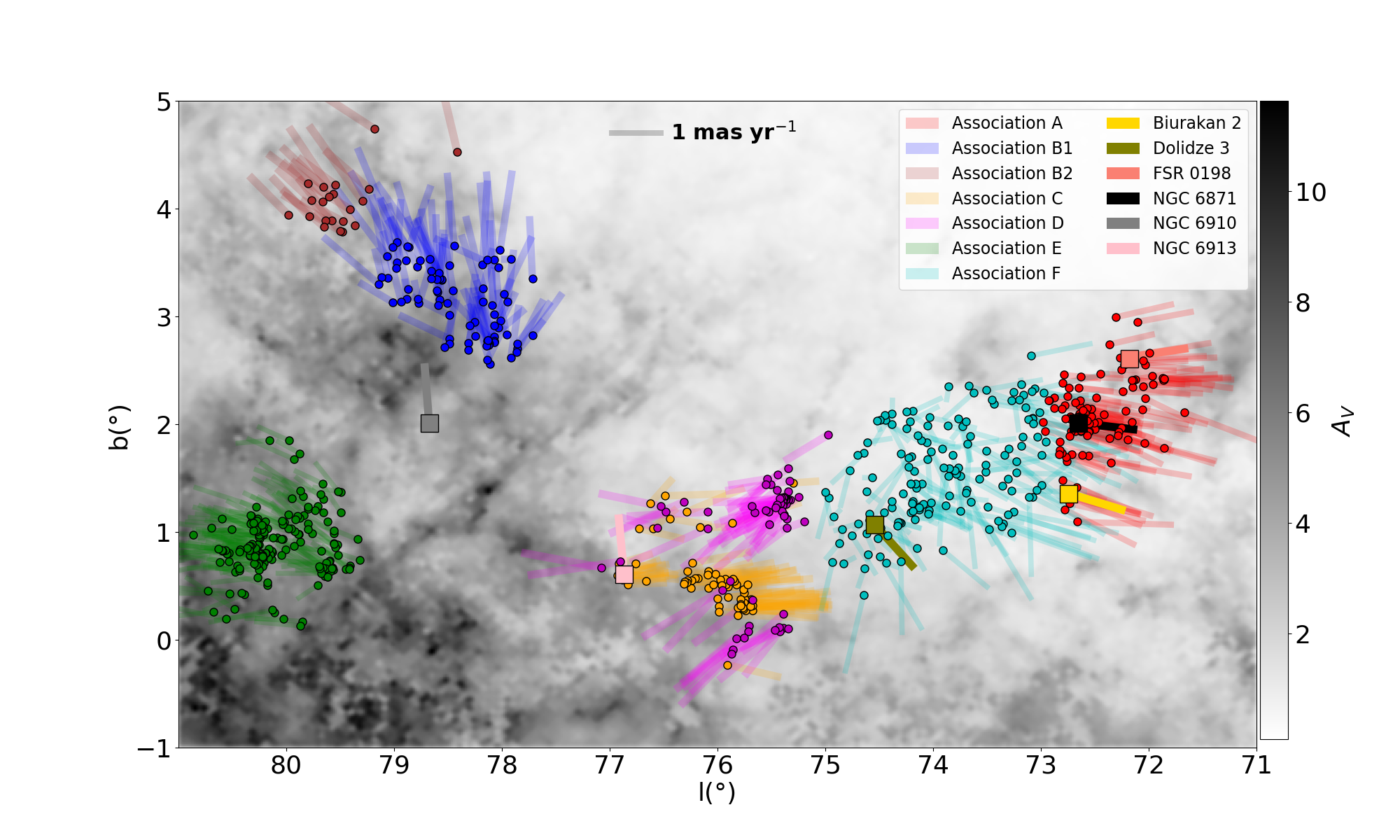}
    \caption{New OB associations in Cygnus identified in \citet{Quintana} and the six selected open clusters in the region, with a background map showing the integrated extinction up to 2 kpc from \citet{Bayestar}. The vectors indicate the PM of each star and open cluster subtracted by the median PMs of all the stars. A representative 1 mas yr$^{-1}$ proper motion vector is shown at the top, equivalent a velocity of $\sim$8.5 km $s^{-1}$ at a distance of 1.8 kpc.
    \label{OBOCs}}
\end{figure*}

The Cygnus region is a well-known region of recent massive star formation \citep[e.g.,][]{Reipurth}, with multiple star forming regions \citep{Schneider2006,Wright2012}, OB associations \citep{Humphreys1978} and star clusters \citep{CantatGaudin}, the most prominent of all of which is Cygnus OB2 \citep{Wright2014, Wright, Wright2016, Berlanas2, Orellana}. In our recent paper, \citet{Quintana}, we studied the known OB associations in Cygnus and found that the majority did not show the kinematic coherence expected for OB associations (Cyg OB2 being the notable exception). We used available photometry and astrometry to identify OB stars across the Cygnus region and identified six new OB associations at distances of 1.5 - 2 kpc, labelled A to F, which we argued should replace many of the existing OB associations in the area. All of these new OB associations are kinematically coherent and each is expanding in at least one dimension. We also discovered a strong correlation between $\mu_l$ and $l$ for the stars in two groups of three associations (ADF and BCE). It was hypothesised that this correlation, indicative of a large-scale expansion pattern, could  be caused by feedback from a previous generation of stars. A similar kinematic pattern was previously observed by \citet{Drew2021} in Carina, who concluded that it could be due to feedback, noting that the scale of the pattern was similar to that observed by \citet{Chevance} in their study of extra-galactic molecular clouds. In this paper we follow up the work of \citet{Quintana} by performing a 3D traceback on the six new associations in Cygnus and to study their past dynamics.

\section{Data}

\label{data}

This section summarises the properties of the OB associations and open clusters in Cygnus and the data used to study them.

\subsection{The new OB associations in Cygnus and recent refinements}

\citet{Quintana} identified six new kinematically-coherent OB associations in Cygnus, replacing some of the previously identified OB associations that were found to be kinematically incoherent. Upon further investigation of association B we noticed a bimodal distribution to its kinematics, and  divided it into two associations, B1 (at lower longitudes and latitudes), containing 79 of the 100 original stars, and B2 (at higher longitudes and latitudes), with 21 stars. Furthermore, refinements to our SED fitting tool found that $\sim$10\% of the OB members of these associations are cooler, A-type stars, but we have retained these stars because they remain kinematic members of the associations.

\subsection{Open clusters}

\citet{CantatGaudin} list 31 open clusters in the Cygnus region studied, including 17 with $d = 1 - 2.5$ kpc. Six of these have PMs and radial velocities (RVs) similar to our associations and therefore are likely to be related. Table \ref{TabOC} lists the properties of these OCs and they are shown in Figure \ref{OBOCs} alongside our 7 new OB associations. Cluster mass estimates for four of the cluster are taken from \citet{Piskunov} but are considered to be conservative as higher estimates exist for NGC 6910 and NGC 6913 \citep{LeDigou}. Dolidze 3 and FSR 0198 lack total mass estimates in the literature and therefore we estimate masses ourselves by fitting the mass function of SED-fitted members from \citet{CantatGaudin} to modelled mass functions from \citet{IMFMasch}.

\begin{table*}
	\centering
	\caption{Properties of the six selected open clusters in Cygnus, with positions, PMs and distances from \citet{CantatGaudin}. The labels for the references for ages, RVs and masses are: K05 = \citet{Kharchenko}, P08 = \citet{Piskunov}, C09 = \citet{Carmago}, D14 = \citet{Dias},  C17 = \citet{Conrad}, C19 = \citet{Carrera}, L19 = \citet{Liu} and K20  = \citet{Kaur}. \label{TabOC}}
	\renewcommand{\arraystretch}{1.3} 
	\begin{tabular}{lcccccccccccr} 
		\hline
		Name & $l(\circ)$ & $b(\circ)$ & $\mu_{\alpha}$ (mas yr$^{-1}$) & $\mu_{\delta}$ (mas yr$^{-1}$) & $d$ (pc) & Age (Myr) & RV (km s$^{-1}$) & Mass ($M_{\odot}$)  \\
		\hline
         NGC 6871 & 72.66 & 2.01 & $-3.13 \pm 0.01$ & $-6.44 \pm 0.01$ & $1841^{+4}_{-3}$ & $7.0 \pm 0.4$ (L19) & $-10.5 \pm 2.2$ (C17) & $1148^{+801}_{-472}$ (P08) \\
         NGC 6910 & 78.68 & 2.01 & $-3.41 \pm 0.01$ & $-5.36 \pm 0.01$ & $1741^{+6}_{-7}$ & $4.25 \pm 1.5$ (L19, K20) & $-32.7 \pm 2.1$ (C17) & $209^{+208}_{-104}$ (P08) \\
         NGC 6913 & 76.87 & 0.61 & $-3.41 \pm 0.01$ & $-5.77 \pm 0.01$ & $1719 \pm 7$ & $5.0 \pm 0.3$ (L19) & $-16.9 \pm 0.6$ (C17)  & $27^{+29}_{-14}$ (P08) \\
         Biurakan 2 & 72.7 & 1.39 & $-3.17 \pm 0.02$ & $-6.84 \pm 0.02$ & $1751 \pm 10$ & 13.8 (K05) & $-22.0 \pm 9.5$ (C17) & $135^{+116}_{-63}$ (P08)\\
         Dolidze 3 & 74.54 & 1.07 & $-2.87 \pm 0.02$ & $-5.61 \pm 0.01$ & $1907 \pm 14$ & $4.0 \pm 0.2$ (L19) & $-7.7 \pm 1.9$ (C19) & $200 \pm 50$\\
         FSR 0198 & 72.18 & 2.61 & $-3.56 \pm 0.03$ & $-6.61 \pm 0.03$ & $1944^{+12}_{-11}$ & $10.0 \pm 5.0$ (C09) & $-13.0 \pm 3.7$ (D14) & $350 \pm 100$ \\
		\hline
	\end{tabular}
\end{table*}

\subsection{Radial velocities}
\label{radialvel}

To obtain 3D kinematics for our associations and open clusters we need RVs to complement Gaia PMs. We gathered RVs from the literature, rejecting those considered unreliable or without measured uncertainties, compiling RVs for 93 stars across our 7 new associations. Since the effects of unresolved close binaries can cause individual measured RVs to vary significantly from the binary system velocity, we calculated the median RV for each association and assigned this to all stars in the association. Unfortunately none of the stars in association B1 have measured RVs, so we took its RV to be the same as that of association B2.

To estimate the uncertainty on the median velocity of each association we perform a Monte Carlo simulation of an association's velocity dispersion to calculate the difference between a system's central velocity and a median derived from N velocities for stars in that association. We assume that each group has a 1D velocity dispersion of 2 $km s^{-1}$ and sample from this distribution. To these velocities we add randomly sampled instantaneous binary offset velocities, assuming a 100\% binary fraction (as observed in Cyg OB2, \citealt{Kiminki2012}), mass ratios from 0.005 to 1.00 with a power-law distribution and an index of 0.1, periods from 1 to 1000 days with a power-law distribution and an index of 0.2, and ellipticities from 0.0001 to 0.9 with a power-law distribution and an index of -0.6 \citep{Kiminki2012}. These velocity offsets are added to the calculated velocities alongside uncertainties sampled from the observed values. We repeat this process 100,000 times and calculate uncertainties from the 16th and 84th percentile values. The results are listed in Table \ref{Radvel} and show that associations with sparsely-sampled RVs have a larger uncertainty on the median RV. 

\begin{table}
	\centering
	\caption{RVs calculated for the new OB associations. $N_{stars}$ is the number of stars in the association whilst $N_{RV}$ is the number of stars with a reliable measured RV. \label{Radvel} The fourth column lists the median RV of the association, and an uncertainty calculated as described in Section \ref{radialvel}. References are: (1): \citet{Hayford}, (2): \citet{Abt}, (3): \citet{Huang2006}, (4): \citet{Gont}, , (5): \citet{Kiminki}, (6): \citet{Huang2010}, (7): \citet{Choj}, (8): \citet{GaiaDR2}, (9): \citet{Holgado2018}, (10): \citet{Carrera}}
	\renewcommand{\arraystretch}{1.3} 
	\begin{tabular}{lccccccccr} 
		\hline
		Assoc. & $N_{stars}$ & $N_{RV}$ & RV (km s$^{-1}$) & References \\
		\hline
        A & 109 & 20 & $-12.50 \pm 2.77$ & (1), (2), (3), (6)  \\
        B1 & 79 & 0 & & \\
        B2 & 21 & 1 & $-21.00 \pm 10.80$ & (4) \\
        C & 75 & 10 & $-19.20 \pm 2.98$ & (1), (3), (6), (7), (10)\\
        D & 65 & 10 & $-7.00 \pm 3.88$ & (1),  (6), (10)\\
        E & 168 & 48 & $-12.35 \pm 2.06$ &  (5), (11) \\
        F & 147 & 4 & $-13.48 \pm 5.06$ &  (3), (8), (10) \\
		\hline
	\end{tabular}
\end{table}

\section{Kinematic traceback}
\label{kin}

In this section we outline the traceback method used to study the past motion of these associations. 

Galactic space velocities $UVW$ were calculated for all stars in our 7 new OB associations and the 6 open clusters themselves (we consider the 6 open clusters as individual objects as they are gravitationally bound, but consider the individual stars in our 7 new OB associations separately since the associations are unbound). These velocities are then corrected for the motion of the Sun $(U_{\odot}, V_{\odot}, W_{\odot}) = (11.10, 12.24, 7.25)$  km s$^{-1}$ from \citealt{Scho}. to obtain quantities relative to the local standard of rest.

\begin{figure}
    \centering
    \includegraphics[scale = 0.10]{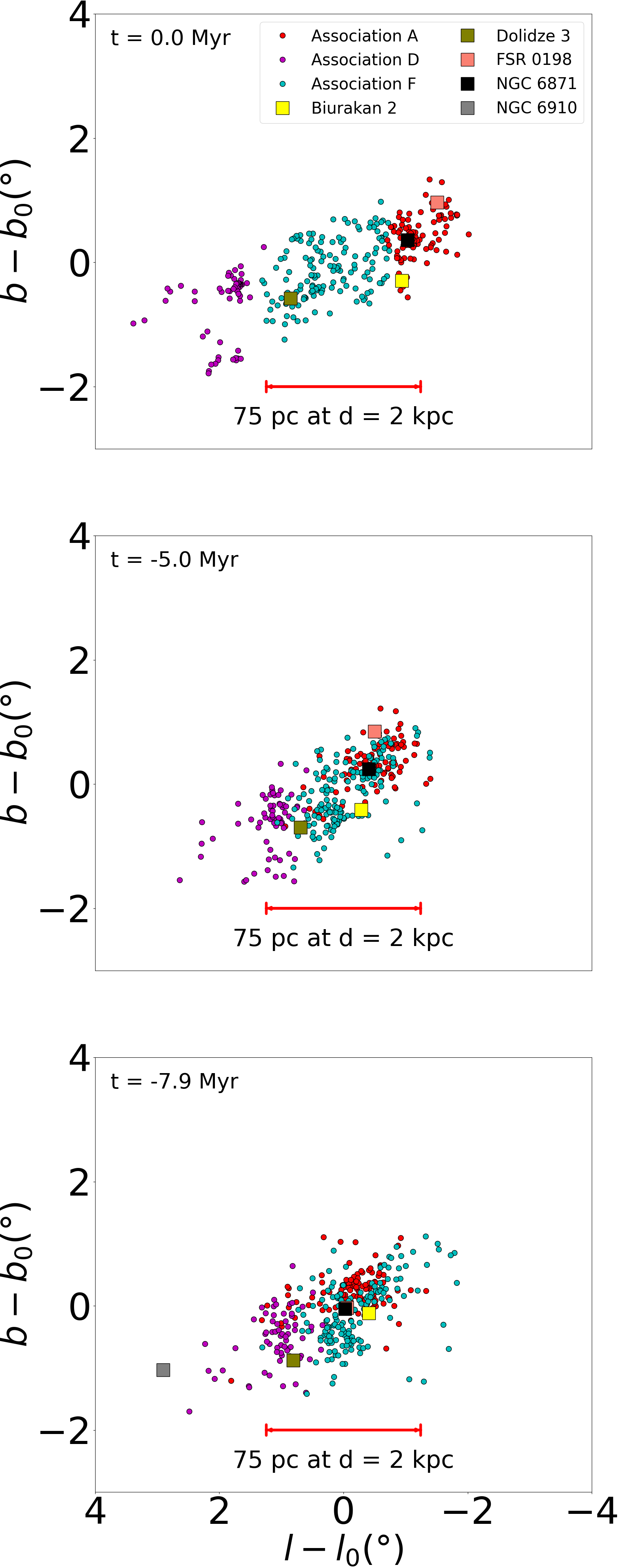}
    \includegraphics[scale = 0.10]{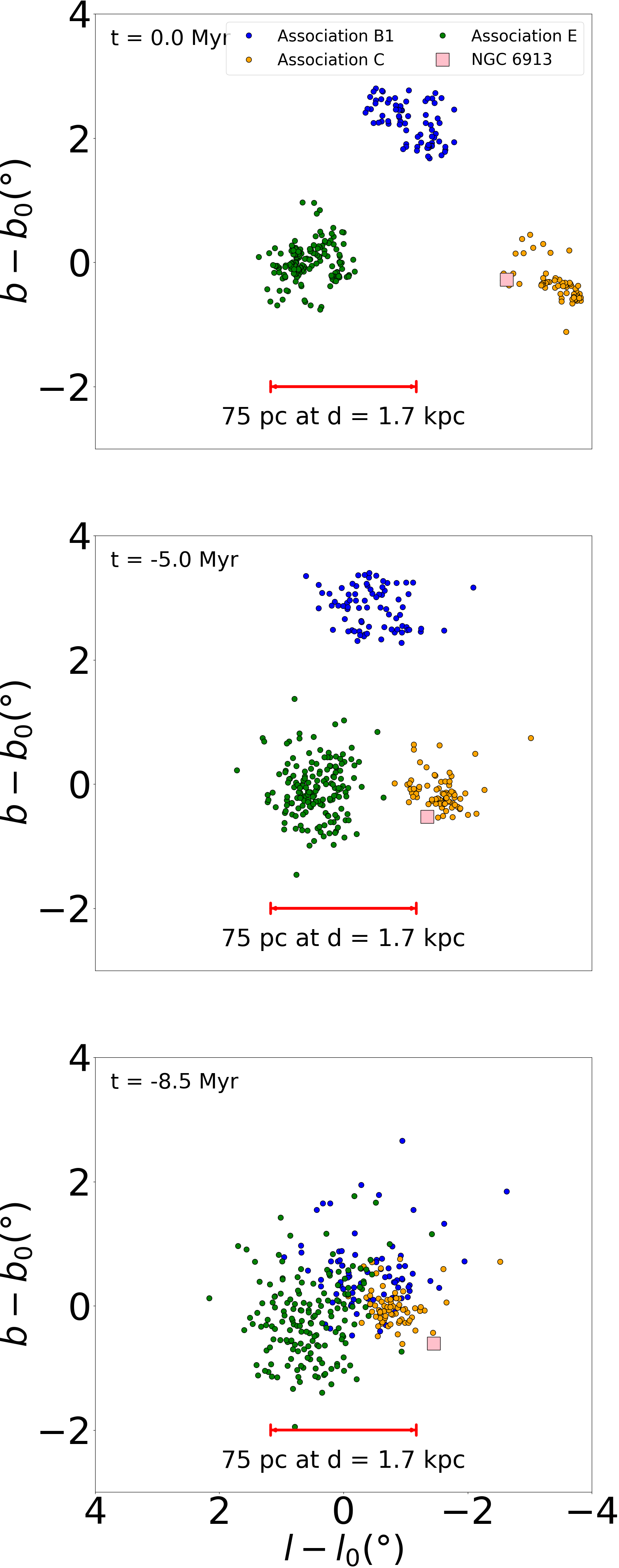}
    \caption{Galactic coordinates of all stars and open clusters at various times in the past derived from traceback calculations in their reference frame, where $l_0$ and $b_0$ stand for the median galactic coordinate of each group.
    \label{Tracegroups}}
\end{figure}

\begin{figure}
    \centering
    \includegraphics[trim={0 0.8 2.8cm 2cm}, clip, scale = 0.3]{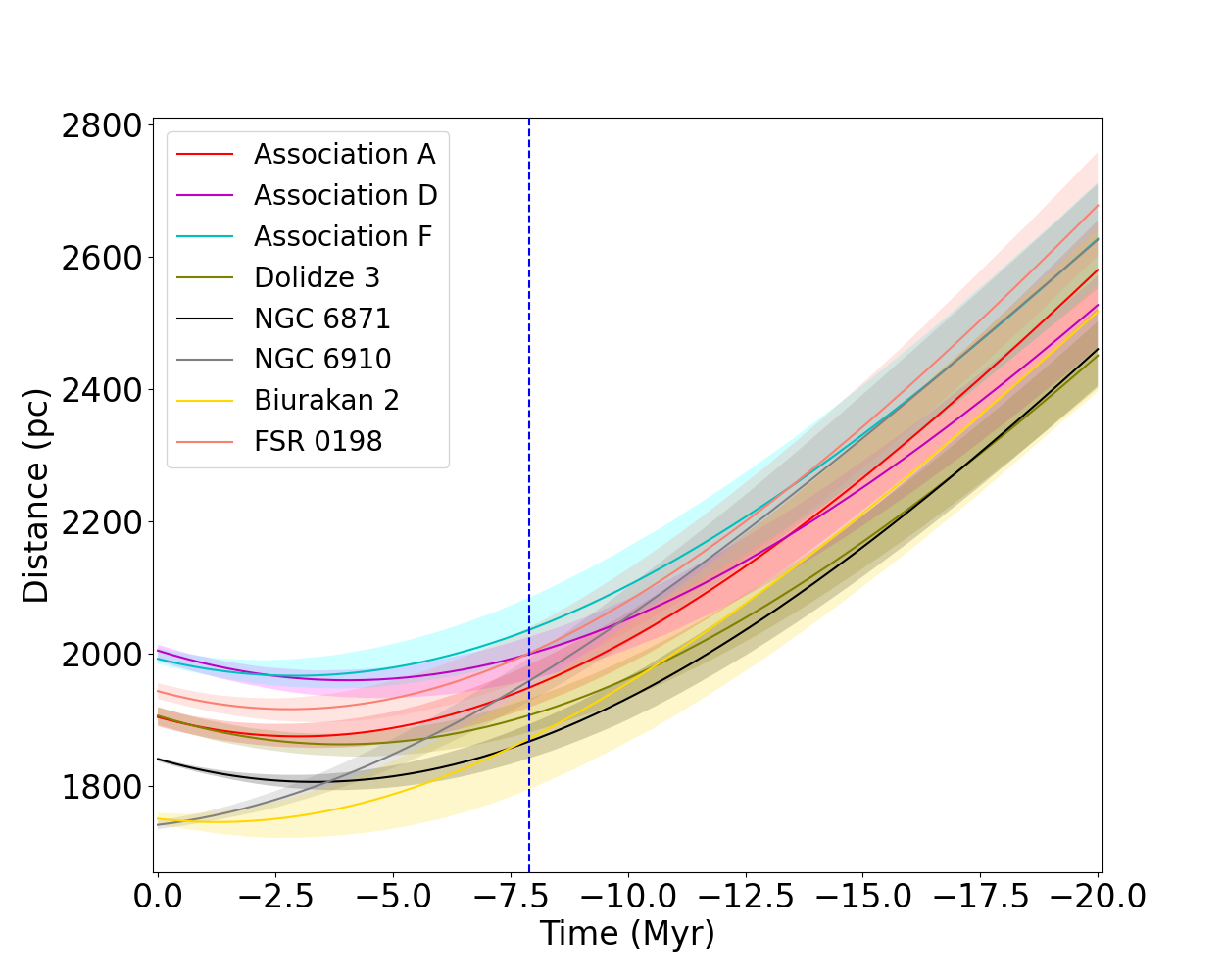}
    \includegraphics[trim={0 0.7cm 2.8cm 2.5cm}, clip, scale = 0.3]{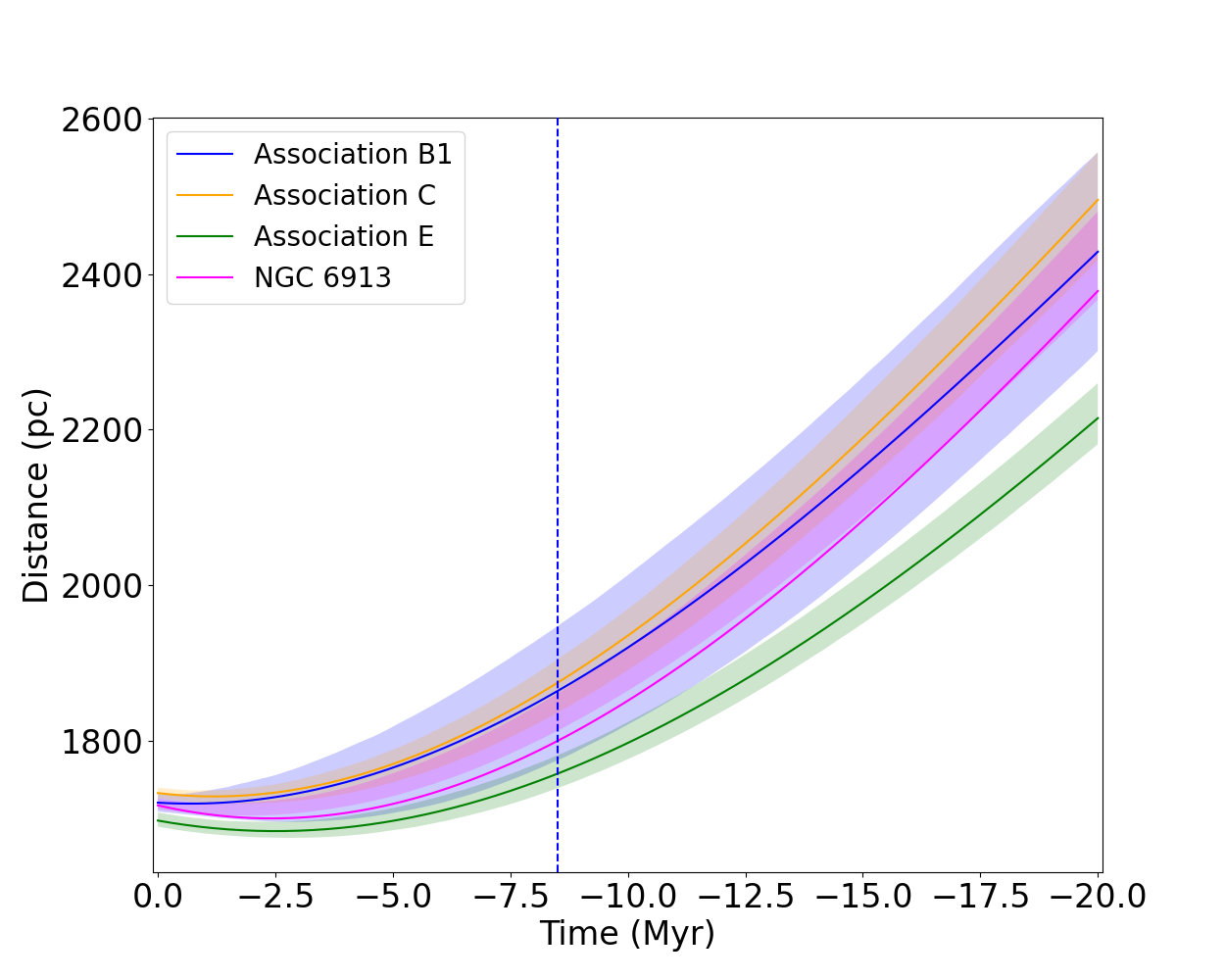}
    \caption{Distance as a function of look-back time for the associations and open clusters studied. Filled areas delimit the 16th and 84th percentiles of distance at each time step. Top panel: associations A, D and F with their related open clusters. Bottom panel: associations B1, C and E with their related open cluster. The dashed blue line shows the time of closest approach on the sky for each group. \label{Dvst}}
\end{figure}

We trace back the motions of the stars and open clusters using the epicycle approximation from \citet{Fuchs}, the Oort constants from \citet{Feast}, and the local density from \citet{Holmberg}. We calculate the $XYZ$ coordinates at times, $t$, up to 20 Myr in the past, at step times of 0.1 Myr. The projected positions of all stars and open clusters are shown as a function of time in Figure \ref{Tracegroups}, while the line of sight distances are shown in Figure \ref{Dvst} (we show the median distance to each association, rather than for individual stars, due to the large uncertainty on individual distances).

As expected from the kinematic signature of expansion identified by \citet{Quintana} the stars in both group ADF and B1CE are closer together on the plane of the sky in the recent past than they are at the current time. This is the case both in the $l$ direction (where the kinematic signature of expansion was identified), but also in the $b$ direction. Associations ADF are also closer together along the line of sight in the recent past, however associations B1CE move slightly further apart in the recent past (though the uncertainties on distance and RV are so large that this is not significant).

We then calculate the distance, $d$, from each star to the central (median) position of each group (either group ADF or group B1CE). Due to the imprecision of the line of sight distances and RVs relative to the plane of the sky positions and velocities, we perform this calculation only in the plane of the sky. For each group of three associations we sum the distances at each time step and determine the time of the most compact configuration as the time this sum is minimised. Uncertainties were estimated using a Monte Carlo process with 1000 iterations, randomly dispersing all measured quantities in each iteration (including all constants used in the epicycle calculations) and using the 16th and 84th percentile values of the resulting time distribution as the upper and lower 1$\sigma$ uncertainties.

The resulting times of most compact configuration are $7.9^{+3.0}_{-1.8}$ Myr for group ADF and $8.5^{+0.8}_{-2.8}$ Myr for group B1CE and the distribution of stars at these times are shown in Fig. \ref{Tracegroups}.

\section{Discussion}
\label{discussion}

We have shown that there are two groups of three OB associations in Cygnus (as well as multiple open clusters) whose kinematics show that they are moving away from each other and that they can be traced back into a more compact configuration in the past. These two groups are at their most compact 7.9 and 8.5 Myr ago, for the associations ADF and B1CE, respectively.  Our calculated traceback ages are generally larger than the evolutionary ages estimated for these associations. \citet{Quintana} estimated that associations B, C and E are 3-5 Myr old while associations A, D and F are older, with stars aged up to  $\sim$10 Myr in them, while the open clusters that are part of these expansion patterns have ages of 5-10 Myr (the exception to this, Biurakan 2, has an age without an uncertainty and therefore its accuracy is unknown). Therefore it would appear that the large-scale expansion we have identified began before these stars and clusters formed and thus the driving force of the expansion would have acted on the primordial molecular clouds and not the stars we observe today. The traceback ages are also consistent with the expansion ages for the individual OB associations estimated by \citet{Quintana, ErratumQuintana}, which are themselves larger than the evolutionary ages for the associations. This suggests that the expansion of the individual OB associations was not driven by their own (stellar) dynamics or processes such as residual gas expulsion, but was seeded prior to their formation, potentially by the same process responsible for the large-scale expansion patterns studied here. A similar chronological pattern is seen in $\lambda$ Ori \citep{Kounkel2018}. 

\citet{Quintana} suggested that the expansion observed could be due to feedback from a previous generation of stars, and the large-scale coherent motions and traceback to a more compact configuration could support this. However, the observed expansion could alternatively be the result of the initial turbulent motion in the primordial molecular cloud \citep{Larson}. We explore both scenarios below.
\subsection{Feedback}
One explanation is that the expansion observed could be due to two major feedback events that occurred $\sim$8 Myr ago and drove the two primordial molecular cloud complexes apart. It is more likely to have been two separate feedback events, rather than a single one, as the two kinematic trends indicative of expansion are distinct (see Figure 11 of \citealt{Quintana}). These feedback events would have swept up the molecular gas in these regions, possibly triggering star formation within them. As noted by \citet{Grob}, such a picture is similar to the classical feedback-driven scenario proposed by \citet{ElmegreenLada1977}, albeit with an emphasis more on 'compress and collapse' rather than 'collect and collapse' as the mechanism for triggering.

To investigate this scenario we estimate how much kinetic energy and momentum is present in the expanding motion of these stars. Since we are only able to observe the stars and yet the energy would have been injected into the primordial molecular clouds, we estimate the mass of these clouds by assuming that they formed stars with a star formation efficiency (SFE) of 5\% and use the association masses estimated by \citet{Quintana} and the open cluster masses from Table \ref{TabOC}. We calculate the velocity of each OB association and OC relative to the mass-weighted mean velocity of each group of associations and clusters and use these to calculate the kinetic energies and momenta injected into each system, assuming they were previously at rest (a highly simplistic assumption). The total kinetic energy injected is $0.54 \times 10^{50}$ erg for group ADF and $1.21 \times 10^{50}$ erg for group B1CE, while the total momentum is $0.66 \times 10^6$ $M_{\odot}$ km s$^{-1}$ for group ADF and $1.26 \times 10^6$ $M_{\odot}$ km s$^{-1}$ for group B1CE.

The total energy output from a supernova explosion is approximately $10^{51}$ erg (see e.g. \citealt{Janka}), a similar level to that calculated for the groups in Cygnus. Predictions from simulations suggest that the total momentum injected into a surrounding molecular cloud by a supernova is between 2 and $4 \times 10^5$ $M_{\odot}$ km s$^{-1}$ \citep{Geen, Kim2015, Walch}, which is between a factor 1.5 and 6 times lower than we have calculated for the Cygnus feedback events. \citet{Kim2015} simulated the feedback generated from ten supernovae and estimated that the final momentum would be in the range of 14 to $22 \times 10^5 M_{\odot}$  km s$^{-1}$, closer to our estimates. This all implies that either several supernovae would have to have contributed to the potential feedback events observed or that the motions observed are due to a combination of feedback types, such as photo-ionization, radiation pressure, stellar winds and supernovae. 

Similar kinematic patterns showing the expansion of large complexes of stars have been observed by \citet{Kounkel2020} and \citet{Grob} in Orion (through respectively the 3D dynamics of stellar groups and the motion of star-forming clouds) and \citet{Drew2021} in Carina (through the motion of OB stars). The large-scale expansion pattern in Orion identified by \citet{Kounkel2020} was attributed to a supernova explosion that occurred approximately 6 Myr ago. \citet{Grob} observed the same expansion pattern in Orion, considered various sources of feedback and concluded that a combination of feedback sources was likely to be responsible. \citet{Grob} estimated the total kinetic energy in the expanding structures in Orion to be (3.5-9.6) $\times 10^{48}$ erg and the total momentum to be (0.7-1.3) $\times 10^5$ $M_{\odot}$ km s$^{-1}$. These estimates are significantly lower than our estimated kinetic energy and momentum, implying that both hypothetical feedback events in Cygnus would have to be considerably more energetic than the Orion event. This is not surprising given that the Cygnus star forming complex is both considerably larger and more massive than the Orion region. 
\subsection{Intrinsic turbulent motions}
Multiple studies have observed a power-law relationship between the physical size of molecular clouds and the 1D velocity dispersion within them, commonly known as Larson's law, which has been attributed to turbulence \citep{Larson, Solomon, Bolatto, Heyer, Miville}. An intrinsic distribution of velocities present in the primordial molecular cloud would presumably be passed onto the stars that form from that cloud. If that velocity dispersion were high and there was not a sufficiently large restoring force, this would be evident as an expansion pattern in the stars that had formed.

To test whether this scenario could explain the observed motions, we can compare the velocity dispersion that Larson's law predicts for an appropriately sized molecular cloud and the velocity dispersion of the expanding stars in Cygnus. Fig. \ref{Tracegroups} suggests a size of $\sim$75~pc at the most compact time for both Cygnus groups. Applying the scaling relations in these papers we find that a molecular cloud with a size of 75 pc would be expected to have a 1D velocity dispersion of 5.7 - 8.7 km/s, depending on the exact relationship used \citep{Larson, Solomon, Miville}. We calculate the actual 3D velocity dispersions using the UVW velocities at the most compact time for both group ADF and B1CE and we respectively find 7.1 and 11.3 km $s^{-1}$, equivalent to 1D velocity dispersions of 4.1 and 6.5 km $s^{-1}$.

From these results, the velocity dispersions of both groups are perfectly consistent with Larson's law. This suggests that the observed expansion of both groups of stars, associations and clusters can be explained as being due to the intrinsic velocity dispersion in the primordial molecular cloud arising from turbulence.  Of course, stellar feedback is thought to be one of the drivers of turbulence \citep{Miville}, and while turbulence may have contributed to the velocity field in the primordial molecular cloud, it appears that feedback is not required to introduce additional momentum to generate the observed expansion pattern.

Repeating this calculation for Orion, Fig. 7 from \citet{Grob} suggests a lower limit of $\sim$40~pc at the most compact time for the cloud size, which implies a 1D velocity dispersion of 4.5-6.3 km $s^{-1}$. We also perform a similar calculation for the Orion subregions in \citet{Grob} (using their UVW velocities in the LSR from their Table 4) and found a 3D velocity dispersion of 4.8 km $s^{-1}$, equivalent to a 1D velocity dispersion of 2.8 km $s^{-1}$. It appears that the observed motions in Orion are also consistent with Larson's law.

\section{Summary}
\label{conclusion}

We have presented the discovery of two distinct expansion patterns in the large-scale distribution of OB stars, associations and open clusters in Cygnus. We have traced back the motion of OB associations and open clusters and shown that they were more compact $\sim$8~Myr in the past as part of two large groups. While it is tempting to attribute this expansion to a specific driving force such as feedback from massive stars or supernovae, we find that the observed kinematics of the expanding systems can be explained as a product of the turbulent velocity field in the primordial molecular cloud. This is not to say that feedback has not had some effect on the dynamics of these stars, but that it is not necessary to search for an additional driving force of the observed motions when they can be fully explained as a result of turbulence. We suggest that a similar conclusion can also be drawn for other recently-observed expansion patterns amongst young stars.

\section*{Acknowledgements}

 We thank the anonymous referee for their careful reading of this manuscript and their useful suggestions. ALQ acknowledges receipt of an STFC postgraduate studentship. This study uses data processed by the Gaia Data Processing and Analysis Consortium (DPAC, https://www.cosmos.esa.int/web/gaia/dpac/consortium) and obtained by the Gaia mission from the European Space Agency (ESA) (https://www.cosmos.esa.int/gaia), along with \textit{TOPCAT} \citep{Topcat}, Astropy \citep{Astropy} and the Vizier and SIMBAD database, both operated at CDS, Strasbourg, France. \\

\section*{Data Availability}
The data underlying this article will be uploaded to Vizier.



\bibliographystyle{mnras}
\bibliography{Main} 




\bsp	
\label{lastpage}
\end{document}